# Beyond Adiabatic Elimination in Topological Floquet Engineering

Yiming Pan[1,*], Ye Yu[2], Huaiqiang Wang[3], Tao Chen[1], Xiaopeng Shen[4,*], Qingqing Cheng[2,*]

[1] Department of Physics of Complex Systems, Weizmann Institute of Science, Rehovot 76100, ISRAEL

[2] School of Optical-Electrical and Computer Engineering, University of Shanghai for Science and Technology, Shanghai 200093, CHINA

[3]National Laboratory of Solid State Microstructures and School of Physics, Nanjing University, Nanjing 210093, CHINA

[4]College of Physical Science and Technology, China University of Mining and Technology, Xuzhou 221116, CHINA

[†]e-mail: yiming.pan@weizmann.ac.il, xpshen@cumt.edu.cn, qqcheng@usst.edu.cn

**Abstract**

In quantum mechanics, adiabatic elimination is a standard tool that produces a low-lying reduced Hamiltonian for a relevant subspace of states, incorporating effects of its coupling to states with much higher energy. Suppose this powerful elimination approach is applied to quasi-energy states in periodically-driven systems, a critical question then arises that the violation of the adiabatic condition caused by driven forces challenges such a presence of spectral reduction in the non-equilibrium driven system. Here, both theoretically and experimentally, we newly reported two kinds of driven-induced eliminations universal in topologically-protected Floquet systems. We named them "quasi-adiabatic elimination" and "high-frequency-limited elimination", in terms of different driven frequencies that deny the underlying requirement for the adiabatic condition. Both two non-adiabatic eliminations are observed in our recently developed microwave Floquet simulator, a programmable test platform composed of periodically-bending ultrathin metallic coupled corrugated waveguides. Through the near-field imaging on our simulator, the mechanisms between the adiabatic and driven-induced eliminations are revealed, indicating the ubiquitous spectral decomposition for tailoring and manipulating Floquet states with quasi-energies. Finally, we hope our findings may open up profound and applicable possibilities for further developing Floquet engineering in periodically-driven systems, ranging from condensed matter physics to photonics.

Adiabatic elimination (AE) is a standard decomposition technique in quantum physics that allows one to get rid of the irrelevant states and produces an effective reduced Hamiltonian for the relevant subspace in closed form. The underlying idea behind adiabatically eliminating certain degrees of freedom out of the dynamics under the previous study has been successfully applied in N-level atomic physics[1-4], quantum optics[3, 5-7], nonlinear optics[5, 8, 9], and plasmonic multi-layer systems[10], but except for periodically-driven systems[11, 12]. Recently, on the other hand, the concept of Floquet engineering in a quantum or classical driven system has gained more and more attention because it enables us to engineer the out-of-equilibrium or synthetic properties of the driven Hamiltonian by steering the drive forces or modulation protocols[11, 12]. However, often the powerful AE technique fails inevitably owing to the violation of adiabatic condition if being extended into a driven circumstance, especially for characterizing of these emergent driven-induced resonant behaviors and non-equilibrium steady dynamics[3, 13, 14].

Floquet topological phases may offer us a remedy. For instance, for a specific driven system, we expect that the Floquet bulk-boundary correspondence, one of the fundamental principles and manifestations in topological insulators and quantum field theories[15-19], can lead to non-adiabatic eliminations in topological driven systems. The correspondence that connects the Floquet quasi-energy band and its isolated protected driven states, indicates the existence of well-separated quasi-energy gaps in Floquet topological insulators. The isolated quasi-energy states are thus topologically-protected by those quasi-energy gaps. In other words, the Floquet gap excludes from the off-resonant bulk states, and thus, produces an effective Hamiltonian for the relevant edge states in a subspace. Consequently, Floquet topological phases enables us to ripe the irrelevant bulk states out, and develop a non-adiabatic elimination for the anomalous edge states in periodically-driven systems.

Here, in this work, we proposed a topology-induced non-adiabatic decomposition technique that can eliminate the irrelevant degrees of freedom of driven systems beyond the restriction of adiabatic theorem. The prominent presence of driven topological states allows us to possibly manipulate and control the quasi-energy spectrum using various drive forces or modulations selectively. Next, as one of the alternative protocols, we can break the degeneracy of those driven edge states with finite-size reduction. By overlapping the edge-state wavefunctions spatially via the hybridization, we can induce a small energy splitting between two degenerated edge states. As a result, we realize a non-adiabatic elimination by tailoring the anomalous Floquet edge states.

With the combination of simulation and experiment, we verified our theoretical expectation and observed the non-adiabatic elimination phenomena in our Floquet simulator composed of the coupled waveguides[20]. With the help of the Floquet simulator, we demonstrated two novel non-adiabatic eliminations in periodically-driven systems, termed as "quasi-adiabatic elimination" (QAE) and "high-frequency-limited elimination" (HFLE). Prospectively, as a promising decomposition technique to quasi-energy band, our newly reported spectral approaches can extend our knowledge of AE to non-equilibrium driven systems, and can be widely applied in current rapidly-developing fields, such as quantum and photonic simulations, ultra-cold atoms and trapped ions, and laser-induced atomic and spin systems[21-26].

**Modeling and setup-** In our simulator, the waveguide arrays are equipped with programmable coupling strengths and non-negligible curving profiles. Through the coupled-mode theory, the driven coupled-waveguides can be accurately mapped into a tight-binding-approximated and time-periodic Schrodinger-like equation ($i\partial\psi/\partial z=\mathcal{H}(z)\psi$)[20], with the guiding propagation direction ($z$) mapping equivalently to the time ($t$). Thus, as a reference, the microwave-propagation of the corresponding Hamiltonian ($\mathcal{H}(z)$) can be expressed as

$$\mathcal{H}(z)-\beta_0\mathbf{I}_{N\times N}=\begin{pmatrix} & \kappa_0-\delta\kappa(z) & & & & \\ \kappa_0-\delta\kappa(z) & & \kappa_0+\delta\kappa(z) & & & \\ & \kappa_0+\delta\kappa(z) & & \ddots & & \\ & & \ddots & & \kappa_0-\delta\kappa(z) & \\ & & & \kappa_0-\delta\kappa(z) & \end{pmatrix}_{N\times N}. \quad (1)$$

where $\beta_0$ is the propagation constant in the weak-guiding approximation, N is the waveguide number, and $\kappa_0$ is the averaged coupling strength between two adjacent waveguides. We assume the staggered coupling term $\delta\kappa(z)=\delta\kappa_0+\delta\kappa_1 cos(2\pi z/\Lambda+\theta_0)$ with the global staggered off-central shift $\delta\kappa_0$, the periodic staggered shift $\delta\kappa_1$, the bending-profiled period $\Lambda$, and the initial position (Floquet gauge) $\theta_0$[11, 12]. If the $\delta\kappa(z)$ is independent of the propagation direction ($z$), these straight coupled waveguides mimic the celebrated Su-Schrieffer-Heeger (SSH) model, which is one of the static one-dimensional topological phases [27].

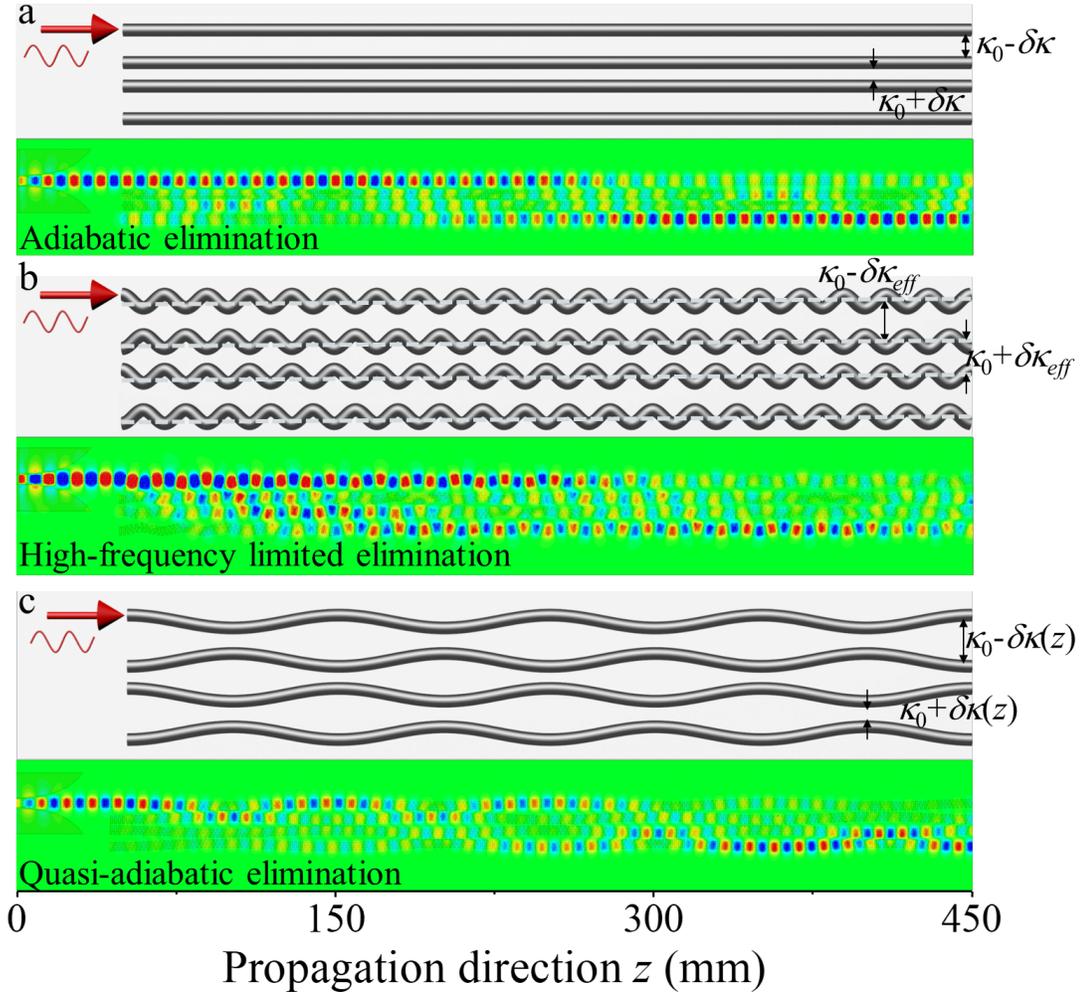

FIG. 1: The coupled-waveguides demonstration of adiabatic elimination (AE), high-frequency-limited elimination (HFLE,) and quasi-adiabatic elimination (QAE) of Floquet engineering. The AE (a) appears at the adiabatic limit ($\Lambda\to\infty$, dimerized array), the HFLE (b) exists at the high-frequency limit ($\Lambda\to 0$, dimerized curved array), and the QAE (c) only emerges at non-trivial driving transition regime in-between the two limits (curved array).

First, for the adiabatic elimination (AE), we consider the straight arrays only consists of four waveguides N=4, in which the staggered coupling $\delta\kappa(z)=\delta\kappa_0$ is constant. Also, the decomposition condition for the slowly-varying guiding approximation has been investigated, which is $|(\kappa_0-\delta\kappa_0)/(\kappa_0+\delta\kappa_0)|\leq 1$, see the derivation in the SM file. As a result, the four-level waveguide array can be reduced adiabatically into the effective two-level subsystem. As shown in Fig. 1a, the waveguide array is simulated with CST commercial software. The static evolution of the electric component ($E_y$, perpendicular to the ultrathin array) shows that the two outer waveguides are coupled effectively by excluding the two inner waveguides out of the field propagation. The array configuration parameters $\kappa_0$=0.03 mm$^{-1}$ and $\delta\kappa_0$=0.02 mm$^{-1}$ confirm the satisfaction of the adiabatic condition $|(\kappa_0-\delta\kappa_0)/(\kappa_0+\delta\kappa_0)| \approx 0.35 < 1$.

Second, beyond the adiabatic condition in periodically-driven systems, the generic driven-induced eliminations are classified in three Floquet-engineered regimes[11, 12, 20, 28]: AE in the adiabatic regime (at driven frequency ω→0), HFLE in the high-frequency regime (at ω→∞), and QAE in the intermediate driven frequency transition regime between the two limits. Notice that the frequency condition of QAE is comparable to the static energy bandwidth of $H_0$ (ω~Δ). The numerical near-field evolutions of HFLE and QAE are shown in Fig.1b and 1c, in which both the propagating patterns appear to be that of AE in Fig. 1a. Their mechanisms, however, are completely distinct from the adiabatic case, because the dynamics of the Floquet simulator is driven into the non-adiabatic resonance regimes, in which the fast-driving protocols has played the predominant role[11, 12]. Given the bending array configurations, we introduce an adiabatic-like condition for quasi-energy spectral decomposition, which is given by

$$\left|\frac{\kappa_0 - \delta\kappa(z)}{\kappa_0 + \delta\kappa(z)}\right| \leq 1, \quad (2)\textbf{Error! Bookmark not defined.}$$

The coupling $\delta\kappa(z)$ implies the influence of the driving frequency ($2\pi/\Lambda$). Suppose the adiabatic-like condition (2) always satisfies, for instance, at $\kappa_0$>>max{$\delta\kappa(z)$} and $\delta\kappa_0$>> $\delta\kappa_1$, thus the quasi-energy decomposition is achieved similar to AE condition. One way to slightly loosen the strong statement (2) associated with universal high-frequency behaviors, is to consider the equivalent static Hamiltonian in high-frequency driven regimes[11, 12].

With the high-frequency approximation, the coupling strength $\delta\kappa(z)$ can be replaced by the effective stroboscopic coupling $\Delta\kappa_{eff} = \lim_{\Lambda\to 0}\int_0^\Lambda \delta\kappa(z)dz$, resulting expectedly into the effective elimination condition $|(\kappa_0-\Delta\kappa_{eff})/(\kappa_0+\Delta\kappa_{eff})|\approx 0.57<1$. Correspondingly, the simulation parameters for HFLE are optimally setup $\kappa_0$=0.029 mm$^{-1}$, $\delta\kappa_0$=0.008 mm$^{-1}$, $\delta\kappa_1$=0.013 mm$^{-1}$, as shown in Fig. 1b.

However, the rigorous argument of condition (2) is exceptionally severe that hardly reflects the resonant stroboscopic behaviors. For the modest driven frequencies ($\omega\sim\Delta$), we found an unexpected eliminated stroboscopic evolution in the frequency range with array parameters $\kappa_0$=0.03 mm$^{-1}$, $\delta\kappa_1$=0.02 mm$^{-1}$, as shown in Fig. 1c. The flow of the input microwave signal propagates along with the two waveguides on the boundary but periodically-driven by virtue of the curving profiles and eventually couples to the two waveguides on the other boundary, roughly analogous to the pattern of adiabatic elimination. We called it "quasi-adiabatic elimination" (QAE), primarily because this driven-induced elimination arises at certain properly non-adiabatic driven frequency regime, close to slowly-driven and resonant frequencies, whose stroboscopic dynamics involves the driven force preparation and setup initialization.

Compared to the high-frequency approximation condition, the QAE does not only violate the strong argument (2) with the staggered coupling strength $\delta\kappa(z)=\delta\kappa\,cos(2\pi z/\Lambda+\theta_0)$ changing its sign varying sinuously, but also it fails in the effective adiabatic-like condition at the high-frequency approximation (i.e., $|(\kappa_0-\Delta\kappa_{eff})/(\kappa_0+\Delta\kappa_{eff})|=1$, $\Delta\kappa_{eff}=0$). Because of lacking a sufficient non-driven rationale compared to the adiabatic condition, we are forced to develop a new decomposition mechanism for elucidating QAE behavior.

The quasi-energy spectrum is one of point of penetration and breach to explain the quasi-adiabatic elimination. Note that the stroboscopic evaluation in the driven system can be decomposed into Floquet quasi-energy states, where the driving frequency drastically changes the spectral structure of quasi-energy band. To further confirm, we calculate its quasi-energy spectrum for QAE in terms of the driving frequencies defined as $\omega=2\pi/\Lambda$ for two cases N=80 and N=4, as shown in Fig. 2a, and 2b.

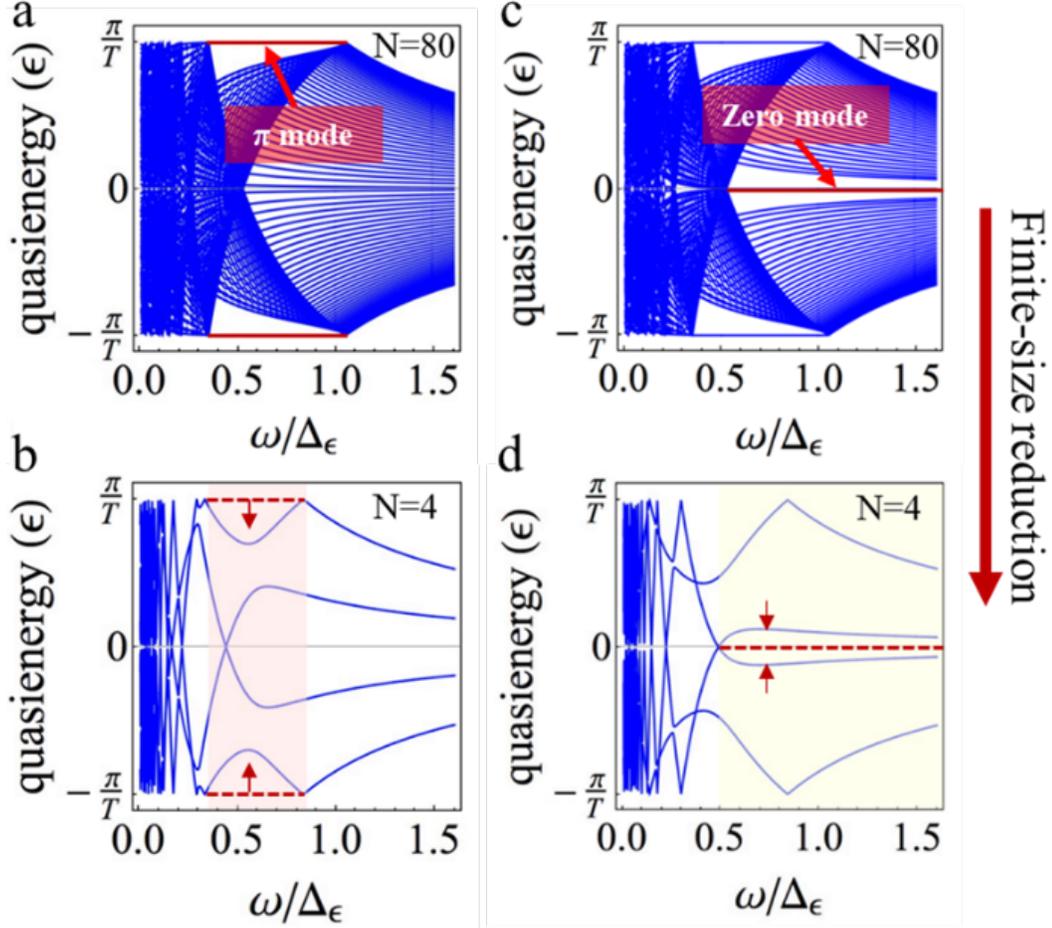

FIG. 2: The physical origins of quasi-adiabatic elimination (QAE) and high-frequency-limited elimination (HFLE) from the violation of degeneracy of π and zero-modes of topologically protected periodically-driven systems. Two typical cases with waveguide number N=80 and N=4 is compared. The quasi-adiabatic eliminations are expected at the conditions $1/3<\omega/\Delta<1$ and $\omega/\Delta>1/2$, respectively.

At the case N=80, corresponding to the periodically-driven Su-Schrieffer-Heeger (SSH) model[27, 29-31], this driven dimerized model has been previously proposed to hold the anomalous topological phase, that is, π modes, which were recently observed in our Floquet simulators[20, 29, 30]. The driven-induced π modes are isolated edge states that appear at degenerate energy positions ($\pm\pi/\Lambda$) in the spectrum (Fig. 2a), associated with driven frequency condition ($1/3<\omega/\Delta<1$) and well-separated from the Floquet bulk states. The topological-protection is associated with the emergence of driven-induced anomalous topological invariant in the quasi-energy gap. Therefore, these protected edge states are driven-dependent, and results in the QAE between two spatially-separated π modes. As compared with the case N=4, the resulting QAE are essential to its survival with violation of its π-mode degeneracy by finite-size hybridization.

**Results and discussions-** In our setup, the waveguide arrays are illustrated numerically and experimentally by using our photonic Floquet simulators composed of four coupled microwave waveguides (N=4) (also see the theoretical results in Fig. S1 of SM file). The waveguides are fabricated using ultrathin metallic coupled corrugated strips, which support Spoof Surface Plasmon polaritons (SPPs)[32, 33], as single-waveguide-mode excitations at microwave frequency ranging from 10 GHz to 18 GHz. The guiding Spoof SPP excitation allows fabrication flexibility and detection visualization to configure a versatile Floquet simulation platform with high propagation and confinement efficiency[20, 34].

Finite-size reduction (e.g., decreasing waveguide number from N=80 to 4) is one of the most straightforward procedure to break the degeneracy and yield an energy splitting between the anomalous π modes. We reduce the waveguide number to four and plotted the quasi-energy band in Fig. 2b, showing the π modes band splitting as compared with the spectrum in Fig. 2a. The simulations are performed in our microwave arrays for the waveguide number ranging from N=8, 6 to 4 in the supplementary Fig. S2, which shows the π-mode field propagation transition from isolation to hybridization. Now each hybridized band has the mixed components of two π modes spectrally and offers them overlap remotely in spatial dimension. While in the condition that this scale of energy splitting does not exceed the π-gap, it remains the protection of QAE behavior. Strictly speaking, the size-induced splitting indeed violates the π modes degeneracy and destroys its topological invariant. Nevertheless, once the size-induced violation is controllably smaller than the driven-induced π gap, the hybridization between two π modes barely scatters into the Floquet bulk, and thus as expected, retains the exclusion of the undesired bulk states.

To evaluate the QAE, we observed the near-field distributions by ejecting the initial microwave signal with different microwave frequencies from 18 GHz to 10 GHz. Note that the microwave frequency is not the Floquet driven frequency defined by waveguide-bending period. As shown in Fig. 3, the QAE field evolution feature vanishes remarkably by decreasing the input microwave frequency into 12 GHz. The practical reason is that the spoof SPP guiding mode has not been well-excited experimentally by low-frequency microwave[20], and the coupling configuration of the array for mimicking time-dependent Schrodinger equation cannot be established. That is to say, our Floquet simulator has to be reconfigured for the given input microwave frequency below 12 GHz.

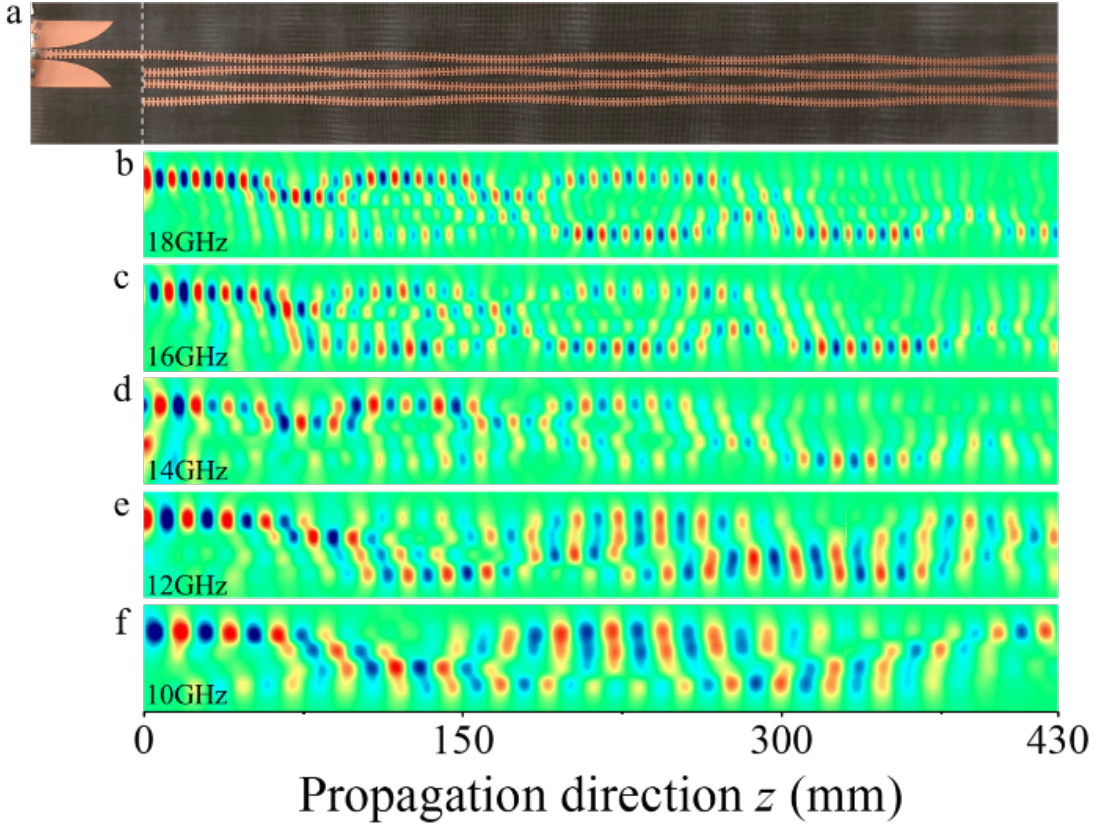

FIG. 3 The diffraction pattern of the Floquet quasi-adiabatic elimination at different frequency from 10 GHz to 18 GHz with the step 2 GHz. (a) the sample of spoof SPP coupled waveguide. (b) quasi-adiabatic elimination $n_\Lambda=3$, $L=430$ mm with the excited frequencies at 18 GHz, (c) 16 GHz, (d) 14 GHz, (e) 12 GHz and (f) 10 GHz.

One intriguing feature of QAE is Floquet gauge dependence[11, 12], as we observed with our fabricated samples and near-field detections, as shown in Fig. 4, and also in Fig. S5 of the SM file. The Floquet gauge is associated with the initial coupling configuration at the array port, in which the gauge is defined as $\theta_0=2\pi z_0/\Lambda$. The parameters of waveguide array profiles follow the simulation parameters (Fig. 1). To test the effect of Floquet gauges, we study four cases in terms of different initial waveguide inputs. As shown in Fig. 4, the QAE stroboscopic patterns are well presented for the two cases (4a) and (4c), but the input microwave fields spread diffusively for the other two cases (4b) and (4d). Explicitly, for the initial field input from the first outer waveguide, the QAE pattern arises at Floquet gauge $\theta_0=0$ but disappears at $\theta_0=\pi$. On the other hand, the second waveguide is initially excited (Fig. 4c), the field evolution at gauge $\theta_0=\pi$ also shows the QAE pattern but no elimination feature at $\theta_0=0$.

This gauge dependence stems from the intrinsic chiral symmetry in our Floquet simulator. Owing to the implicit chiral symmetry $\mathcal{H}(0,-\delta\kappa_1)= \mathcal{H}(\Lambda/2, \delta\kappa_1)$ in the array (Eq. 1), we assert that the stroboscopic dynamics with input from the first waveguide at Floquet gauge $\theta_0=0$ is equivalent to that from the second waveguide at $\theta_0=\pi$. In general, the QAE is dependent subject to driven modulations with the bending profiles (both the period and the gauge). The characteristic of frequency and gauge dependence promises QAE to be potentially engaged into novel optical switches or modulators, and applicable to non-adiabatic driven pumping and transport of light flow. Besides, we notice that the HFLE is universal gauge-independent that have been widely studied in the high-frequency regime[11,12].

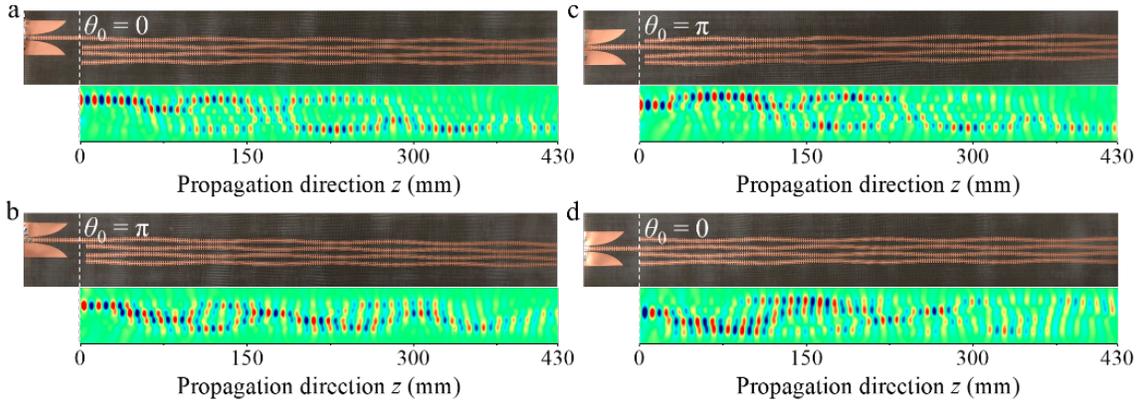

FIG. 4 Near-field measurements of Floquet gauge dependence of quasi-adiabatic elimination (QAE). The equivalence between input positions and Floquet gauge dependence from the symmetric analysis combined the sublattice and time-reverse symmetries with the case (a) equivalent to (c) as well as the case (b) to (d). The white scale bar is 5cm.

Notably, we address a quasi-adiabatic transition between two opposite limits in Floquet engineering, especially unveiling the failure of the adiabatic theorem in this driven-induced phenomena. Instead, in this resonantly driven situation, one must confront the intricate quasi-energy spectrum of Floquet systems. The driving frequency and Floquet gauge in the resonantly driven regime would yield many intriguing stroboscopic dynamics and behaviors as a by-product retreat. Beyond adiabatic elimination in periodically driven systems brings us a new research direction, dubbed as "Floquet photonics". As a promising implementation, it can be further engaged in the optical device design and photonic integration.

To extend, we propose a perspective for application and engineering with anomalous topological phases, as shown in Fig. 5. Many newly-published Floquet-engineered results would fall into three relevant driven protocol regimes, including universal fluctuations[28] and quantum/adiabatic pumping (4D QHE)[19, 35, 36] in the adiabatic regime, anomalous topological phases[17, 20, 37, 38], Floquet gauge dependence[11, 12, 20] and Floquet phase transitions[17, 37, 39] in the quasi-adiabatic transition regime, and universe high-frequency behaviors[11, 12, 17, 22, 40-44] in the high-frequency regime. Our non-adiabatic eliminations to decompose quasi-energy spectrum could be influential in exploring exotic stroboscopic phases of matter, and out-of-equilibrium driven behaviors.

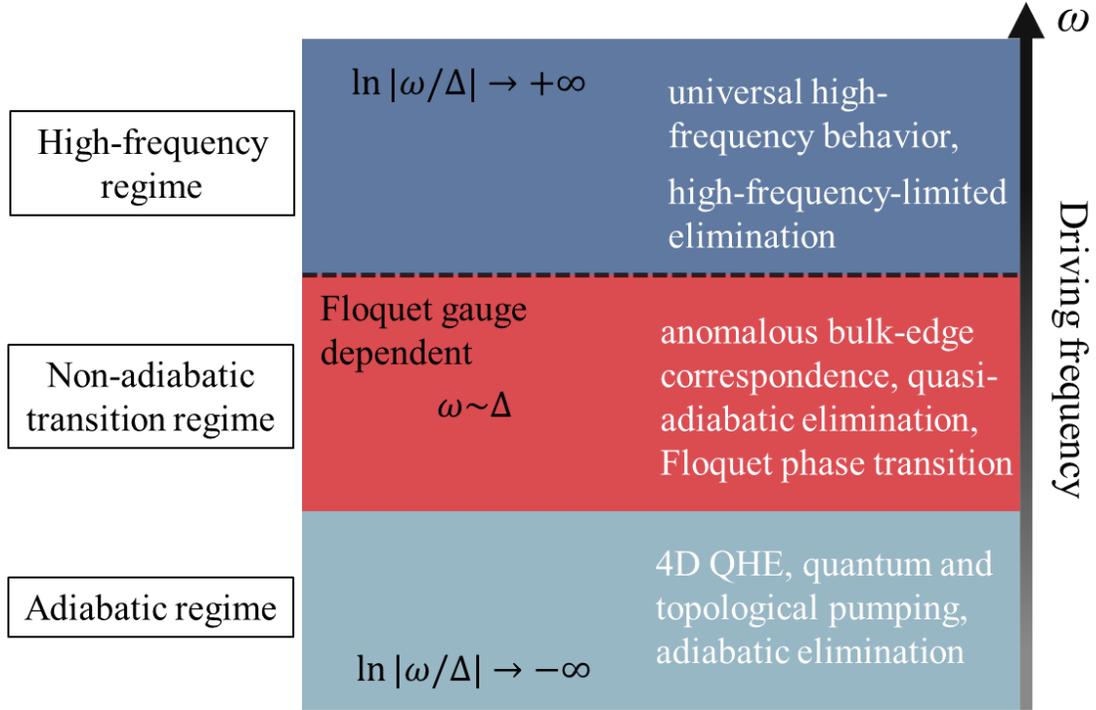

Fig. 5: Illustration of Floquet photonics in fully driven frequency regimes as classified by our developed three types of adiabatic and non-adiabatic eliminations in periodically-driven systems. Two important factors, the driven frequency and the Floquet gauge in Floquet engineering are addressed. Several related works from recent literature are partly reviewed, see the text.

**Conclusions:** We reported two driven-induced eliminations, QAE and HFLE, for the exclusion of irrelevant degrees of freedom in Floquet systems, because the adiabatic condition is explicitly violated due to the drive forces or modulations. We adapted the generalized bulk-boundary correspondence to identify and tailor the topologically-protected edge states from the quasi-energy spectrum, and realize the non-adiabatic eliminations by finite-size hybridization. Significantly, the non-adiabatic elimination holds unique controllable features that depend on the driving frequency and Floquet gauges. The mechanisms of non-adiabatic eliminations can steer the topological periodically-driven systems, and as a powerful and compelling tool, of analyzing the quasi-energy spectrum and its driven-induced dynamics in the fruitful Floquet engineering.

**Method: Sample fabrication and simulation.** The sample is fabricated in a printed circuit board (F4BK) with dielectric constant 2.65, loss tangent 0.001, and thickness 0.2 mm, the copper film thickness 0.018 mm. Taking account for eliminating reflections in the experiment, we add an absorbing material with length thirty millimeters coating the end of the waveguide lattices.

**Experimental measurement.** We conduct the near-field measurements using a near-electric-field platform composed of a Vector Network Analyzer (VNA) (E5063A), a monopole antenna as the detector, and a translation stages that can move in the x and y-directions automatically controlled by a stepper motor. The plasmonic waveguide's input port is connected to port 1 of the VNA through the SMA to feed the energy, and the end is adhered to the absorber to eliminate reflections. To probe the vertical (z) components of the electric fields, the monopole antenna is fixed on top of the plasmonic waveguide around 1.4 mm and connected to port 2 of the VNA for recording the propagation pattern.

**Funding:** This work is supported by the National Natural Science Foundation of China (11874266, 11604208), Shanghai Science and Technology Committee (16ZR1445600), ChenGuang Program (17CG49), and by DIP—German-Israeli Project Cooperation (7123560301), by the BSF-NSF (2014719), by Icore— Israel Center of Research Excellence program of the ISF, and by the Crown Photonics Center, and by the PBC Programme of the Israel Council of Higher Education.

**Acknowledgements:** The authors thank Yaron Silberberg, Haim Suchowski, Ady Arie, Avraham Gover for their helpful discussions and comments. Supplementary information is available in the online version of the paper. Correspondence and requests for materials should be addressed to Y. P., X. P. and Q. C..

# Supplementary Materials for "Beyond Adiabatic Elimination in Topological Floquet Engineering"

Yiming Pan[1],*, Ye Yu[2], Huaiqiang Wang[3], Tao Chen[2], Xiaopeng Shen[4,*], Qingqing Cheng[2,*]

1. Department of Physics of Complex Systems, Weizmann Institute of Science, Rehovot 76100, ISRAEL
2. Shanghai Key Lab of Modern Optical System and Engineering Research Centre of Optical Instrument and System (Ministry of Education), University of Shanghai for Science and Technology, Shanghai 200093, CHINA
3. National Laboratory of Solid State Microstructures and School of Physics, Nanjing University, Nanjing 210093, CHINA
4. College of Physical Science and Technology, China University of Mining and Technology, Xuzhou 221116, CHINA

†e-mail: yiming.pan@weizmann.ac.il, xpshen@cumt.edu.cn, qqcheng@usst.edu.cn

# Contents

## 1. Theory of Adiabatic Elimination

For our concern, consider the Schrodinger equation as a dynamical system, evolving with a Hamiltonian H as $i\partial_t\psi = \mathcal{H}\psi$ (we set $\hbar=1$). Let state $\psi$ be partitioned into $\alpha=\mathcal{P}\psi$ and $\gamma=\mathcal{Q}\psi$, with $\mathcal{P}$ and $\mathcal{Q}=1-\mathcal{P}$ projector, in such a manner that the eigenvalues of $\mathcal{PHP}$ are widely separated from those of $\mathcal{QHQ}$. Let it be that case in which the coupling between the $\mathcal{P}$ and $\mathcal{Q}$ subspaces is very small when compared to the eigenvalues of $\mathcal{QHQ}$. To be more specific, then, the Schrodinger equation may be rewritten as

$$\begin{aligned} i\partial_t\alpha &= \mathcal{PHP}\alpha + \mathcal{PHQ}\gamma \\ i\partial_t\gamma &= \mathcal{QHP}\alpha + \mathcal{QHQ}\gamma \end{aligned} \quad (1)$$

where $\mathcal{P}^2=\mathcal{P}$, $\mathcal{Q}^2=\mathcal{Q}$. Heuristically, we are asking that if $\gamma$ is small and if we neglect its time evaluation $\partial_t\gamma\approx 0$, and in other words, we slave $\gamma$ to $\alpha$ in the adiabatic approximation $\mathcal{QHP}\alpha+\mathcal{QHQ}\gamma=0$, or formally, $\gamma=-(\mathcal{QHQ})^{-1}\mathcal{QHP}\alpha$. By substituting this approximation in the first component of the equations, we obtain the effective evolution for the component $\alpha$ as

$$i\partial_t\alpha = \mathcal{PHP}\alpha - \mathcal{PHQ}\frac{1}{\mathcal{QHQ}}\mathcal{QHP}\alpha = \mathcal{H}_{eff}\alpha \quad (2)$$

By eliminating the irrelevant component $\gamma$, we obtain an effective Hamiltonian for the relevant degrees of freedom that we are interested in

$$H_{eff} = \mathcal{PHP} - \mathcal{PHQ}\frac{1}{\mathcal{QHQ}}\mathcal{QHP} \quad (3)$$

As comparison, we apply the AE decomposition with the propagation of the monochromatic microwave fields in four coupled straight waveguides, as shown in

Fig.2a. The standard coupled mode equations is given by

$$i\frac{d}{dz}\begin{pmatrix}\psi_1\\\psi_2\\\psi_3\\\psi_4\end{pmatrix}=\begin{pmatrix}\beta_0 & \kappa_1 & 0 & 0\\\kappa_1 & \beta_0 & \kappa_2 & 0\\0 & \kappa_2 & \beta_0 & \kappa_1\\0 & 0 & \kappa_1 & \beta_0\end{pmatrix}\begin{pmatrix}\psi_1\\\psi_2\\\psi_3\\\psi_4\end{pmatrix} \quad (4)$$

where $z$ is the propagation direction, and $\psi_i$, $i=1, 2, 3, 4$ represent the amplitude of waveguide mode $i$. All the waveguide modes have some propagation constant $\beta_0$ and the coupling strengths of two adjacent waveguides are characterized by $\kappa_1$ and $\kappa_2$. This configuration is the minimal size (waveguide number $N=4$) of the famous Su-Schrieffer-Heeger model, and one could predict the two zero-modes isolating in the band gap at $\kappa_1<\kappa_2$. Thus, we will isolate the two outer waveguide modes ($|\psi_1>$, $|\psi_4>$) from the others ($|\psi_2>$, $|\psi_3>$), and produce the effective Hamiltonian for the two zero modes. Firstly, we introduce a gauge transformation to get rid of the constant propagation constant, which is,

$$\psi = \phi e^{-i\beta_0 z} \quad (5)$$

Substituting into eq.4, one obtains

$$i\frac{d}{dz}\begin{pmatrix}\phi_1\\\phi_2\\\phi_3\\\phi_4\end{pmatrix}=\begin{pmatrix}0 & \kappa_1 & 0 & 0\\\kappa_1 & 0 & \kappa_2 & 0\\0 & \kappa_2 & 0 & \kappa_1\\0 & 0 & \kappa_1 & 0\end{pmatrix}\begin{pmatrix}\phi_1\\\phi_2\\\phi_3\\\phi_4\end{pmatrix} \quad (6)$$

Secondly, we define the projector $\mathcal{P}$ and $\mathcal{Q}$ which are given by

$$\mathcal{P}=\begin{pmatrix}1 & 0 & 0 & 0\\0 & 0 & 0 & 0\\0 & 0 & 0 & 0\\0 & 0 & 0 & 1\end{pmatrix}, \quad \mathcal{Q}=1-\mathcal{P}=\begin{pmatrix}0 & 0 & 0 & 0\\0 & 1 & 0 & 0\\0 & 0 & 1 & 0\\0 & 0 & 0 & 0\end{pmatrix} \quad (7)$$

Applying into the formula (eq.3), we get the effective Hamiltonian

$$\mathcal{H}_{eff}=\mathcal{P}\mathcal{H}\mathcal{P}-\mathcal{P}\mathcal{H}\mathcal{Q}\frac{1}{\mathcal{Q}\mathcal{H}\mathcal{Q}}\mathcal{Q}\mathcal{H}\mathcal{P}=\begin{pmatrix}0 & 0 & 0 & -\frac{\kappa_1^2}{\kappa_2}\\0 & 0 & 0 & 0\\0 & 0 & 0 & 0\\-\frac{\kappa_1^2}{\kappa_2} & 0 & 0 & 0\end{pmatrix} \Rightarrow \mathcal{H}_{eff}=\begin{pmatrix}0 & -\frac{\kappa_1^2}{\kappa_2}\\-\frac{\kappa_1^2}{\kappa_2} & 0\end{pmatrix} \quad (8)$$

where the basis spans with ($\phi_1$, $\phi_4$). The condition for the decomposition is given by

$$\left\|(\mathcal{Q}\mathcal{H}\mathcal{Q})^{-1}\mathcal{Q}\mathcal{H}\mathcal{P}\right\| \approx 1 \Rightarrow \left|\kappa_2^{-1}\kappa_1\right| \approx 1 \Rightarrow \left|\kappa_1\right| < \left|\kappa_2\right| \tag{9}$$

In general, for N coupled waveguides array, to get rid of the N-2 waveguides ($\phi_2$, $\phi_3$, …. $\phi_{N-1}$) and isolate the two outer waveguide modes ($\phi_1$, $\phi_N$), we could get the form of the effective 'two-level' Hamiltonian as like eq.8 but the coefficient is exponential decay with N, which is,

$$\left\|\mathcal{H}_{\mathit{eff}}\right\| \propto \exp(-N/N_{cr}) \tag{10}$$

where the critical waveguide number is dependent on the amplitude of gap, i.e., $N_{cr}=N_{cr}(|\kappa_1-\kappa_2|/\|\kappa_1+\kappa_2\|)$ Here compare eq.6 and eq.8, following the adiabaticity of AE, the rapidly varying variables of the inter waveguide modes ($\phi_2$, $\phi_3$, …. $\phi_{N-1}$) must slave the adiabatically show varying variables of two outer modes ($\phi_1$, $\phi_N$) and this are thus eliminated.

## 2. Experiment and simulation: Floquet non-adiabatic elimination

2.1 Fig. S1: Numerical results for Floquet engineering non-adiabatic elimination

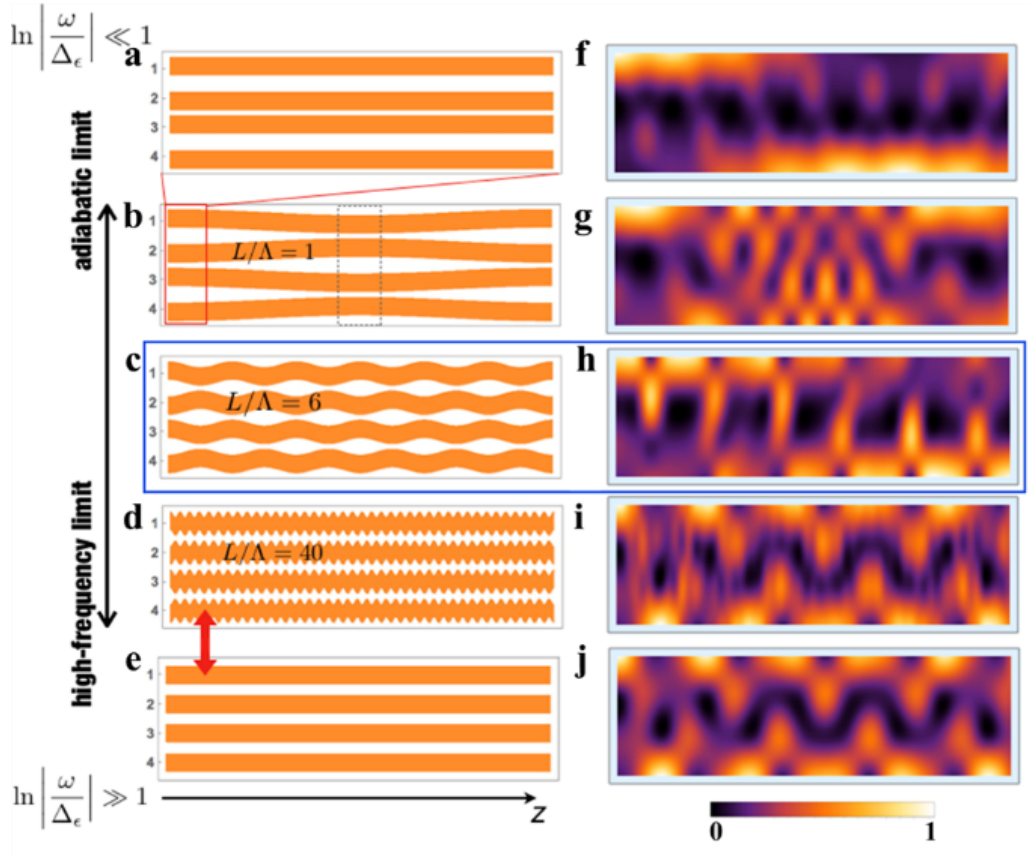

Fig.S1: (a) structure and (f) numerical results for traditional adiabatic elimination, (b) structure and (g) numerical results for modulation adiabatic elimination, (c) structure and (h) numerical results for non-adiabatic elimination, (d) structure and (i) numerical results for High-frequency comparison with (e) structure and (j) numerical results for uniform waveguide arrays.

## 2.2 Fig. S2: The finite size effects in Floquet non-adiabatic elimination

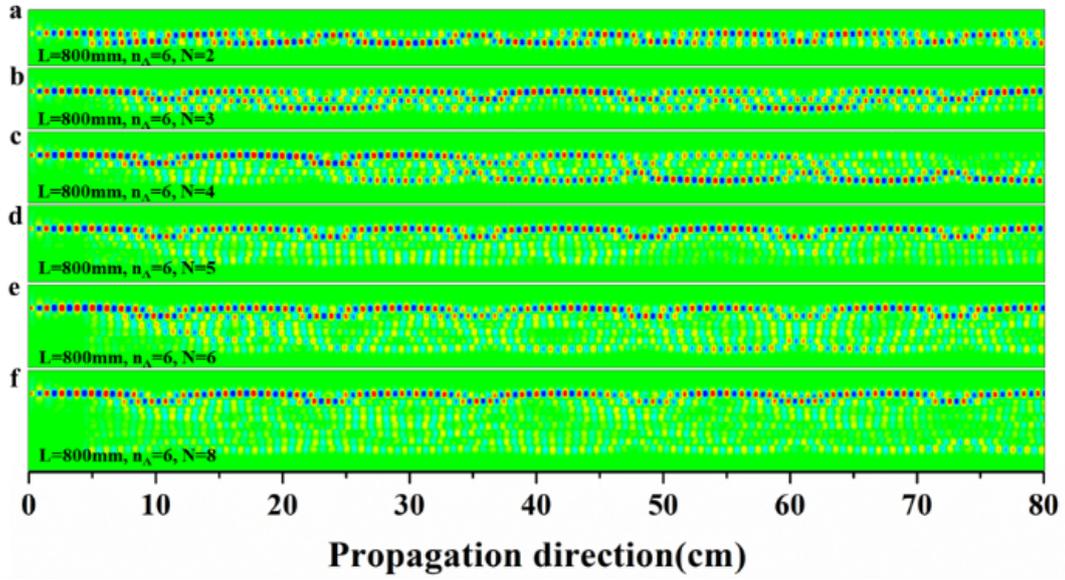

Fig. S2: The diffraction pattern with consideration on $n_\Lambda$ and waveguide number N at the same total waveguide length L=80 cm. The array configuration parameters are $\kappa_0$=0.042, $\delta\kappa$=0.02 for the non-adiabatic elimination.

## 2.2 Fig. S3: Determine the offset between the boundary and the inner waveguide

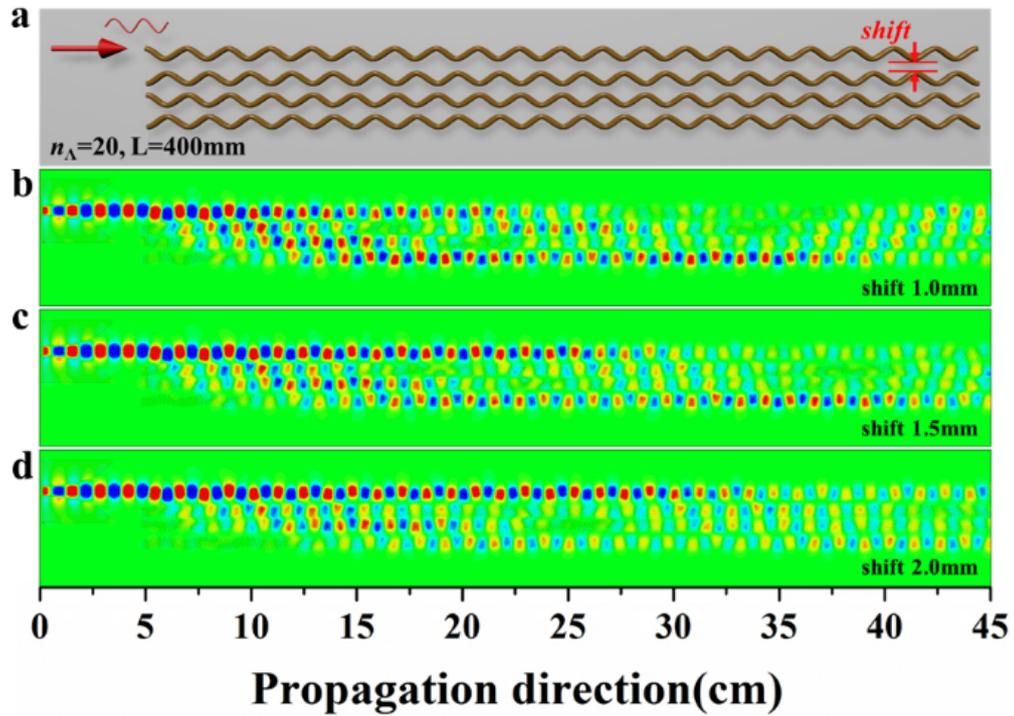

Fig. S3: (a) the configuration of high frequency waveguide arrays with the propagation direction periodic number $n_\Lambda$=20, the total length L=45 cm, the diffraction patterns of the high-frequency waveguide arrays with varying the offset of the boundary and the inner waveguides in (b) the offset 1.0mm, (c) 1.5m and (d) 2.0mm. Consideration of the finite propagation length, we choose the offset at 1.5mm in our experiment.

## 2.4 Fig. S4: Experimental results of the adiabatic and high-frequency elimination

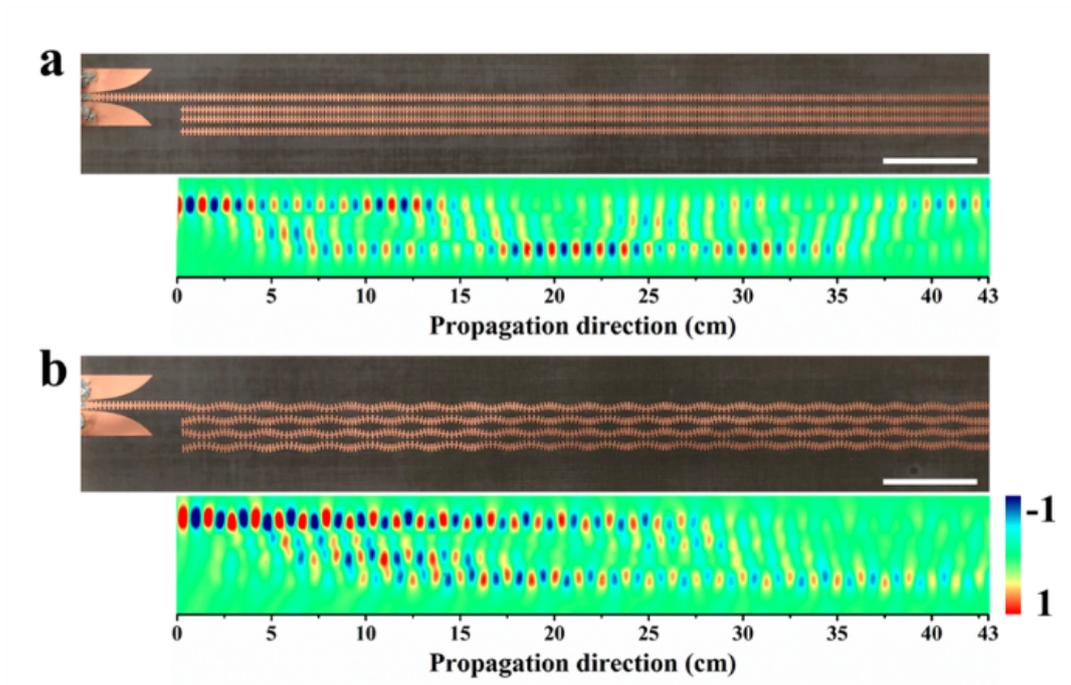

Fig. S4 Near-field measurements of adiabatic elimination (a) and high-frequency-limited elimination (b) in two opposite driving limits. The array configuration parameters are $\kappa_0$=0.042, $\delta\kappa$=0.02 for the adiabatic elimination. The parameters are optimally setup $\kappa_0$=0.042, $\delta_0$=0.042, $\delta\kappa$=0.02 for the high-frequency elimination.

## 2.5 Fig. S5: Floquet gauge dependence of non-adiabatic elimination

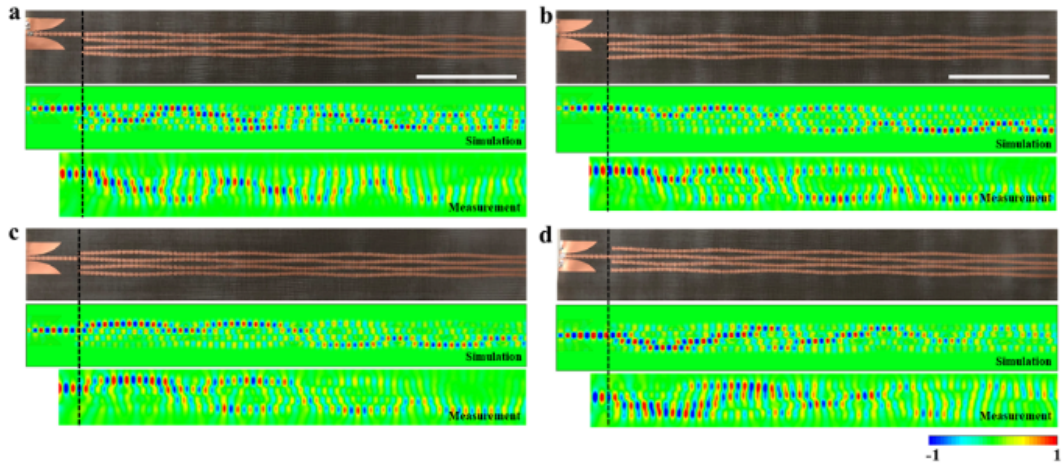

Fig. S5: The equivalence between input positions and Floquet gauge dependence from the symmetric analysis (combined the sublattice and time-reverse symmetries). (a-d) Four different combined Floquet gauges and input positions situations. The case (b) is equivalent to (c) and both of them share a same driven QAE channel, which indicated the potential waveguide integrations. On the other hand, the cases (c) and (d) are trivial. The white length scale is 10 cm.

## 2.6 Fig. S6: Relation of Floquet non-adiabatic elimination with the incoming microwave frequency

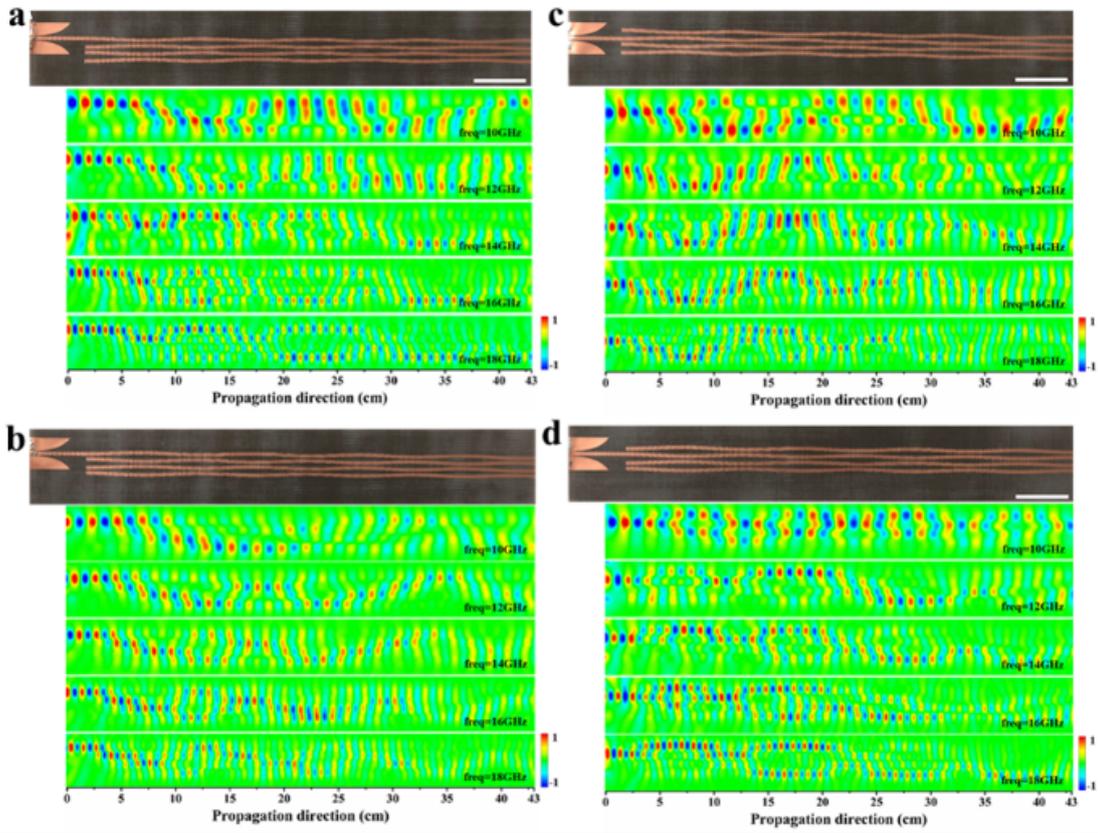

Fig. S6: The diffraction pattern of the Floquet non-adiabatic elimination at different frequency from 10GHz to 18GHz with the step 2GHz. (a) Non-adiabatic elimination $n_\Lambda=3$ L=43 cm with the excited waveguide sequence number 1 and the original phase 180deg, (b) Trivial diffraction pattern with the excited waveguide sequence number 1 and the original phase 0deg, (c) Trivial diffraction pattern with the excited waveguide sequence number 2 and the original phase 0deg, (d) Non-adiabatic elimination $n_\Lambda=3$ and L=43 cm with the excited waveguide sequence number 1 and the original phase 0deg.